\documentclass{Interspeech2024}
\interspeechcameraready 
\usepackage{subcaption}
\usepackage{amsmath}
\usepackage{xcolor}
\usepackage{verbatim}
\usepackage{array}
\usepackage[space]{cite}
\usepackage{amsmath,amssymb,amsfonts}
\usepackage{algorithm}
\usepackage{algorithmic}
\usepackage{textcomp}
\usepackage{graphicx}
\usepackage{xcolor}
\usepackage{tabularx}
\usepackage{textcomp}
\usepackage{multirow}
\usepackage{multicol}
\usepackage{comment}
\usepackage{amssymb}
\usepackage{url}
\usepackage{float}
\usepackage{authblk}
\usepackage{etoolbox}
\usepackage{adjustbox}
\usepackage{pgf}
\usepackage{soul}
\apptocmd{\thebibliography}{\small}{}{}
\patchcmd{\thebibliography}{\leftmargin\labelwidth}{\leftmargin\labelwidth\itemsep=0pt\parsep=0pt\topsep=0pt}{}{}

\title{NeuRO: An Application for Code-Switched Autism Detection in Children}
\name[affiliation={1}]{Mohd Mujtaba}{Akhtar*}
\name[affiliation={1}]{Girish*}{}
\name[affiliation={2}]{Orchid} {Chetia Phukan*}
\name[affiliation={3}]{Muskaan}{Singh*}
\address{
  $^1$School of Computer Science, UPES, Dehradun, India, $^2$IIIT-Delhi, India,  \\ 
  $^3$CARL, ISRC, SCEI, Ulster University\\
   *equal contribution}
\email{mmakhtar.research@gmail.com, girish.research.pr@gmail.com, orchidp@iiitd.ac.in, m.singh@ulster.ac.uk}

\begin{document}

\maketitle

\begin{abstract}
Code-switching is a common communication phenomenon where individuals alternate between two or more languages or linguistic styles within a single conversation. Autism Spectrum Disorder (ASD) is a developmental disorder posing challenges in social interaction, communication, and repetitive behaviors. Detecting ASD in individuals with code-switch scenario presents unique challenges. In this paper, we address this problem by building an application \textbf{NeuRO} which aims to detect potential signs of autism in code-switched conversations, facilitating early intervention and support for individuals with ASD. 
\end{abstract}
%\vspace{-0.5cm}
\section{Introduction}
Autism Spectrum Disorder (ASD), is a complex neuro-developmental disorder due to synaptic dysfunction and altered brain connectivity. It affects an individual's daily life with aspects such as social interaction, communication, and behavior. These aspects emerge in early childhood, as early as 3 years and vary in degrees and their symptoms range from mild to severe. Children with ASD often struggle with basic social cues, such as making eye contact, responding to their names, sharing interests, and moving with others.
Early diagnosis and intervention are crucial for children with autism, as they can receive speech and language therapy, occupational therapy, and social skills training to support them and their families. %\cite{daniolou2022efficacy}.  
Previous studies focusing on the diagnosis of audio focused on low-level acoustic characteristics such as pitch, intensity, and rhythmic pattern as potential markers\cite{DBLP:conf/acl-clpsych/TanakaSNTN14}.
%%DBLP:conf/interspeech/ChoLRCSP19}. 
In addition to these characteristics of vocal outburst and utterance level, which contribute to understanding of audios\cite{cho19_interspeech,eni23_interspeech}, some studies investigated the incorporation of para-lingual features\cite{DBLP:conf/mlsp/ShahinASDE19}, such as voice quality, articulation and resonance, improving the overall performance of automatic autism detection. %However, these hand-crafted features pose limitations \cite{DBLP:conf/eacl/ProbolM24}, which led to a recent shift towards deep learning architectures such as CNN, RNN and transformer-based models to automatically learn discriminative representations from raw speech signals outperforming the traditional feature-based approach.  
%Most of the previous works on audio-based autism detection have primarily relied on handcrafted features or spectral features, without much emphasis on leveraging speech pre-trained model (PTM) features.
In this work, we explore the potential of incorporating linguistic and paralinguistic features for improved code-switched ASD detection. By experimenting with linguistic and paralinguistic features, we aim to capture more comprehensive and discriminative representations of speech for autism detection and the topmost performance is achieved by CNN classifier with paralinguistic features.

\section{System Design}

\begin{figure}[H]
    \centering
    \includegraphics[scale=0.2]{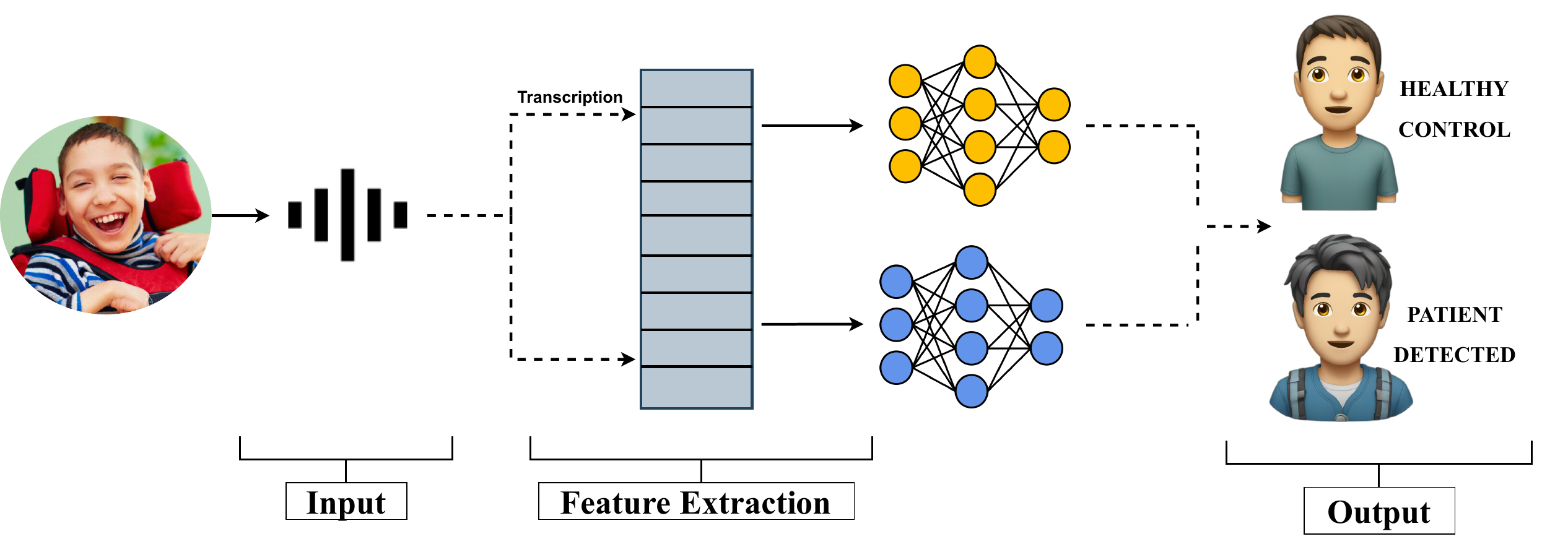}
    \caption{\centering System Design}
    \label{fig:work_flow}
\end{figure}

The modeling system design behind our application, \textbf{NeuRO} is shown in Figure \ref{fig:work_flow} for autism detection. The dataset to be used for modeling, raw audio samples from two distinct groups: children potentially diagnosed with autism spectrum disorder (PT) and healthy controls (HC). These raw audio are further pre-processed by resampling to 16 kHz.
\textit{For linguistic feature extraction}, we employ the Whisper \cite{radford2023robust} model for speech transcription and use the Natural Language Toolkit (NLTK) to extract linguistic characteristics from code-mixed transcripts. These extracted features cover the average lengths of words and sentences, the rate of speech, and the language distribution, providing information on code mixing and language switching patterns between English and Hindi. On the other hand, we extract paralinguistic features using TRILLsson \cite{shor22_interspeech}. \par

We explore classical ML techniques such as SVM, RF and also RNN, CNN, Transformer encoder for experimentation with individual linguistic and paralinguistic features. For SVM and RF, we use the default parameters. For RNN, we use 50 hidden states and the output layer with sigmoid function. We use CNN with 64 filters with size 3, then succeeded by hidden layer with 128 neurons followed by the output. We use transformer with number of heads as 4 with one hidden layer of 128 neurons followed by the output layer with sigmoid function. We also lead an investigation into combining linguistic and paralinguistic features for autism detection. We use only the CNN and transformer models, as it has shown the top most performance with individual models. 
%These linguistic and paralinguistic features are then fused into a combined feature set, which is processed through parallel pipelines comprising 1D-CNN, MaxPooling, and transformer encoder layers. Additionally, we employ machine learning models such as SVM, Random Forest (RF), K-Nearest Neighbors (KNN), Recurrent Neural Networks (RNN), and Transformer Encoder (TE). To establish a baseline, these models are used with their default parameters.vThe cornerstone of our methodology is a \textit{novel CNN model}. This CNN model processes the input voice files through a 1D convolutional layer, followed by a max pooling layer, a transformer encoder, and finally a dense layer with a softmax activation function for classification. 
We train the neural models (RNN, CNN, Transformer) for 50 epochs. We employ 5-fold cross-validation with 4 folds used for training and one fold for testing. %In the final stages, the pre-processed features, as well as the predictions from the machine learning models, are fed into a dense layer for further enhancement. The SoftMax classification is then applied to obtain the final diagnosis of autism spectrum disorder. This comprehensive approach leverages both linguistic and paralinguistic cues, combined with machine learning and deep learning models, to achieve accurate and early detection of ASD, showcasing the potential of our audio-based system for autism diagnosis.
We present the results of the models used in our system design in Table \ref{tab:results}. All of our results are computed on an internally curated dataset of children in code-mixing scenarios alternating between English and Hindi. The data collection comprises interactive sessions soliciting responses from children aged 3 to 13. The total duration of 159.75 minutes in 61 participants was equally divided into 105.6 minutes for 30 ASD and 53.99 minutes for 31 non-ASD. Our data set is diverse and strong, ready for detailed analysis and training classification models, as evident from our results in Table \ref{tab:results}. CNN with paralinguistic features obtained the best performance. We were limited in our experimental setup due to the scarcity of data; to the best of our knowledge, there is no existing code-mixed audio corpus to extend the generalizability of our system design.

%and it is systematically partitioned into ASD and Non ASD classes. %This classification strategy is instrumental in facilitating a nuanced analysis, offering insights into potential patterns or correlations within the dataset.
%It is crafted through a process of soliciting responses from children aged 3 to 13. A specially created original script forms the foundation for interactive sessions with both ASD and non-ASD children and ensures that our dataset is diverse and strong, ready for detailed analysis and training classification models. For this project, two speech sound databases were created. One database comprises speech sound samples from children diagnosed with ASD, while the other database comprises speech samples from non-ASD children. Both databases were recorded in the English and Hindi languages.

\begin{table}
\centering
%\tiny
\setlength{\tabcolsep}{15pt}
% \scriptsize
\caption{Evaluation Scores; Scores are average of 5-folds; F1 stands for macro-average f1-score}
\label{tab:results}
\begin{tabular}{lcc} \toprule
\textbf{\textcolor{blue}{Model}} & \textbf{\textcolor{blue}{Accuracy}} & \textbf{\textcolor{blue}{F1}} \\ \midrule
\multicolumn{3}{c}{\textbf{\textcolor{red}{Linguistic Representation Modeling}}} \\ \midrule
SVM          &       86.98       &        81.44        \\
RF        &  85.12           &        80.80           \\
%KNN             &   81.40           &  75.27          \\ 
%NB           &    80.47         &  74.19  \\
%DT              &  79.07           &   75.38               \\
RNN             &      85.12        &    78.90         \\ 
Transformer            &  88.37          &  85.14 \\
%Conformer            &         87.90     &  83.50 \\ 
CNN           &         87.44     &  80.49\\ 
 \midrule  
\multicolumn{3}{c}{\textbf{\textcolor{red}{Paralinguistic Representation Modeling}}} \\ \midrule
SVM                &    95.81         & 94.67  \\
RF                 &       94.77      &    93.50         \\ 
%KNN                  &        97.57      &    97.11      \\ 
%NB                 &      92.34        &     90.41       \\
%DT              &        89.72     &      87.18            \\
RNN             &        83.72      &     76.82        \\ 
Transformer            &  97.38          &      96.84        \\
%Conformer            &          97.67     & 96.50   \\
CNN            &    \textbf{98.13}    & \textbf{97.37}\\ \midrule
\multicolumn{3}{c}{\textbf{\textcolor{red}{Fusion with Linguistic+Paralinguistic}}} \\ \midrule

Transformer            &       97.81     & \textbf{97.37}  \\ 
%Conformer            &      98.13         & 98.00  \\ 
CNN           &  96.27    & 95.16 \\ 
\bottomrule
\end{tabular}
\end{table}

\begin{figure}[H]
    \centering
    \begin{subfigure}[b]{0.45\textwidth}
        \centering
        \includegraphics[scale=0.28]{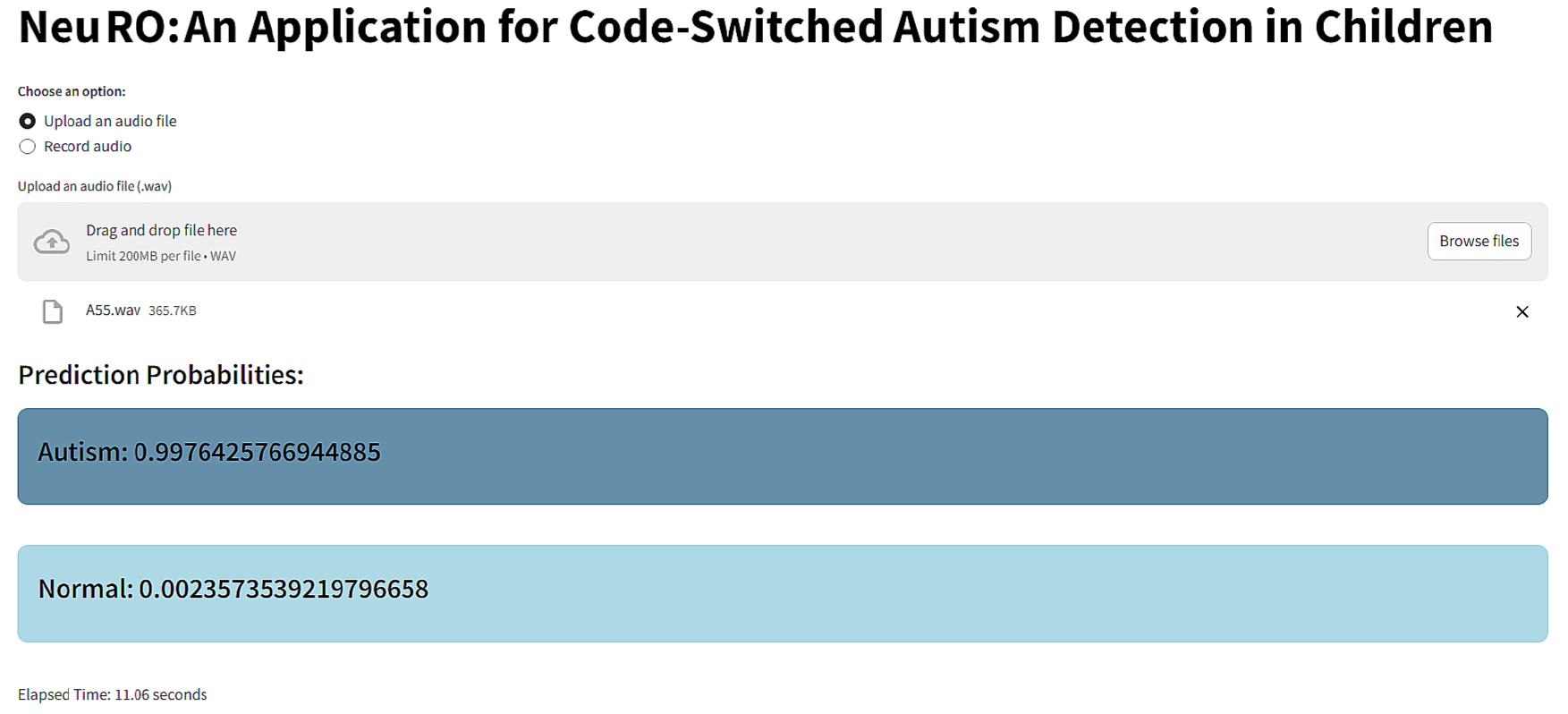}
        \caption{}
        %\label{fig:main-interface}
    \end{subfigure}
    \hfill 
    \begin{subfigure}[b]{0.45\textwidth}
        \centering
        \includegraphics[scale=0.28]{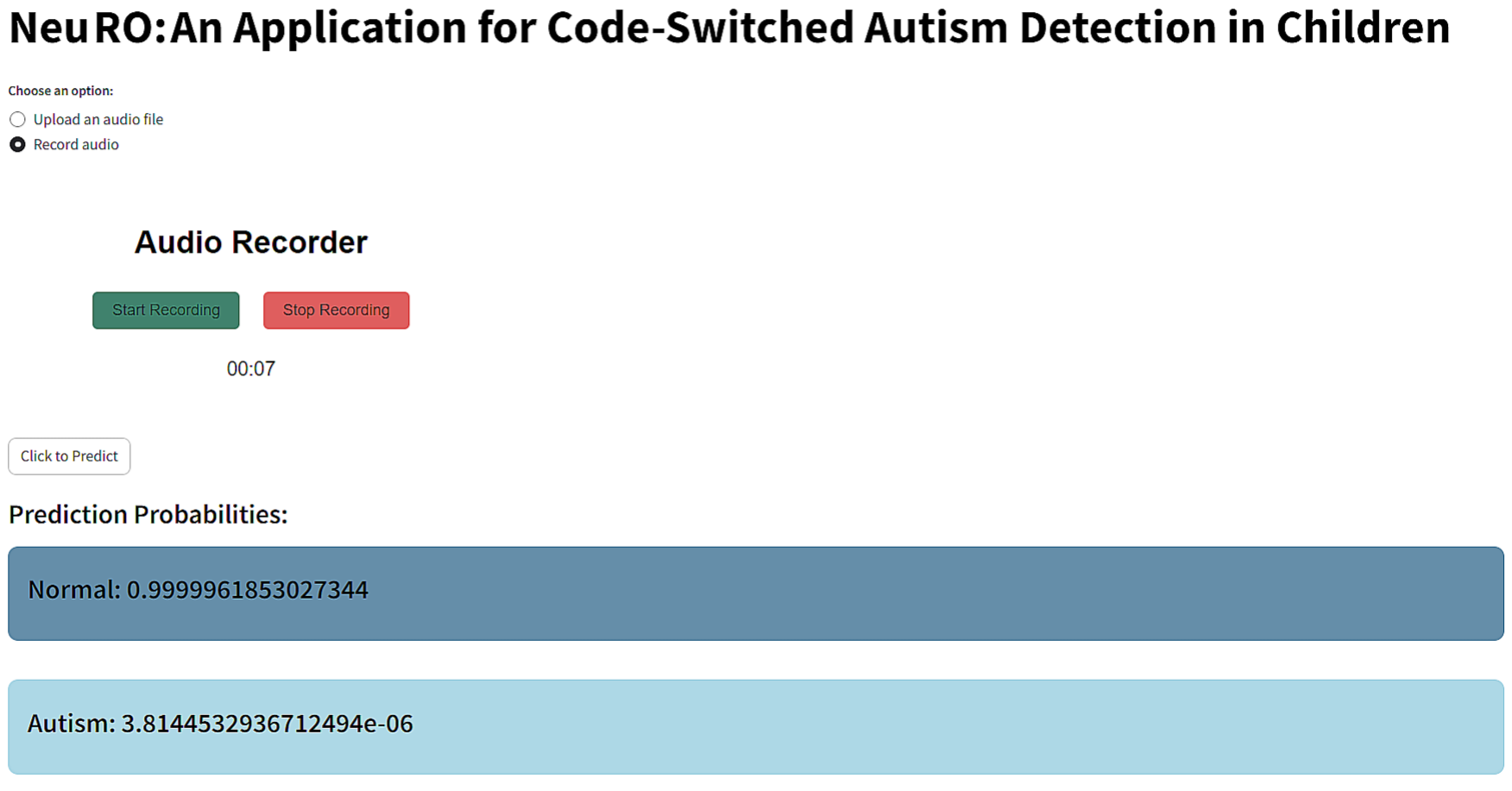}
        \caption{}
        %\label{fig:audio-record-interface}

\end{subfigure}
    \caption{NeuRO Application Interfaces: (a) Audio from Device (b) Real-time Audio Recording }
    \label{fig:neuro-interfaces}
\end{figure}

\section{NeuRO Interface}
To enhance the accessibility and user-friendliness of our autism detection tool, we developed and integrated a web-based user interface, as shown in Figure \ref{fig:neuro-interfaces}. The user interface (UI) was designed to be user-friendly, allowing even non-technical users to easily navigate the platform. After the user uploads a voice recording, the back-end system analyzes the file and extracts features, as described in our approach. Subsequently, the extracted features are then fed into our trained models for classification into PT or HC. This made it easy for data to move from the user interface to the models.
\section{Conclusion}
In this paper, we present a novel real-world application, \textbf{NeuRO}, for the detection of autism from code-switched speech. The model at the core of the system is CNN model with paralinguistic features. This model is seamlessly integrated into the web-based UI to provide a prediction of healthy control or PT from a given input audio. In addition, users can intuitively interact with the system through the system interface provided. Overall, we believe that our proposed system has the potential for widespread usability in diverse fields, such as healthcare, education, and training. 

\bibliographystyle{IEEEtran}
\bibliography{main.bib}

\end{document}